\documentclass[a4paper,12pt]{article}
\usepackage{authblk}
\usepackage[T1]{fontenc}
\usepackage{times}
\usepackage{setspace}
\usepackage[utf8x]{inputenc}
\usepackage[english]{babel}
\usepackage{url}
\usepackage{quoting}
\usepackage{geometry}
\usepackage[hang]{footmisc}
\usepackage{color}

\definecolor{blu}{rgb}{0.12,0.12,0.76}
\definecolor{rosso}{rgb}{0.76,0.12,0.12}

\date{}

\begin{document}

\author[1]{Emilia Margoni}
\author[2]{Daniele Oriti}
\affil[1]{Département de Philosophie, Université de Genève}
\affil[2]{Departamento de Física Teórica, Universidad Complutense de Madrid}

\title{\textbf{The emergence of spacetime: what role for functionalism?}}

\maketitle

\begin{abstract}
\noindent 
Among the various attempts to formulate a theory of quantum gravity, a class of approaches suggests that spacetime, as modeled by general relativity, is destined to fade away. A major issue becomes then to identify which structures may inhabit the more fundamental, non-spatiotemporal environment, as well as to explain the relationship with the higher-level spatiotemporal physics. Recently, it has been suggested that a certain understanding of functionalism is the proper tool to suitably account for the recovery of spacetime. Here the viability and usefulness of such a conceptual strategy is explored, by looking at the various levels of spacetime emergence a theory of quantum gravity is expected to deal with. Our conclusion will be that, while its viability is clear also in a quantum gravity context, the import of spacetime functionalism remains rather unsettled.  
\end{abstract}


\section{Introduction}

The idea according to which the spatiotemporal structure of general relativity (GR) can be regarded as a macroscopic description – analogous to the thermodynamic or hydrodynamic  account of molecular physics – of an underlying set of non-geometric degrees of freedom dates back to as early as the development of black holes thermodynamics in the 1970s (Padmanabhan, 2010, 2014, Carlip, 2014, Oriti, 2014). Indeed, among many other considerations,\footnote{See, e.g., Reid (2001), Henson (2009). For a philosophical discussion see W\"uthruch (2012).} the discovery, in the context of semi-classical gravity, that black holes can be associated with a finite entropy was generally understood as pointing towards the possibility of there being some underlying discrete microstructure of spacetime, given that black holes are but particular spatiotemporal configurations (Bekenstein, 1973).\footnote{On black hole entropy see also Jacobson (1995, 2016).} 

A main task of quantum gravity (QG) becomes then to specify what this microstructure might be, and how this underlying set of degrees of freedom could explain (among other things) black holes thermodynamics. A number of approaches in QG strongly suggest that this putative underlying set of entities cannot be given a spatiotemporal, first, and then gravitational interpretation (at least in the usual language in which space and time are understood in modern physics, i.e. in terms of fields).
Rather, the set of more fundamental degrees of freedom is to be conceived of as an ensemble of possibly discrete, non-geometric entities whose collective behaviour can, under suitable conditions, yet only approximately, recover the usual notions of geometry and gravity in terms of (quantum) fields, including the spacetime metric (Seiberg, 2006, Oriti, 2014, 2017).

In this theoretical scenario, the recovery of spacetime requires both a classical and a continuum approximation. The first is the technically and physically tricky but conceptually straightforward approximation of the quantum behaviour of some physical entity (such as fields or particles) with its corresponding classical one. The second kind of approximation takes different concrete forms depending on the precise nature of the quantum entities one adopts as a starting point, and it may or may not be implemented as a standard continuum limit of, say, lattice theories. As such, the latter has more generally to do with the collective dynamics of a large number of the underlying entities and usually involves some form of coarse-graining and averaging of key quantities. It is especially with respect to the latter approximation that the conceptual framework of certain QG approaches gets closer to the formalisms and methodologies of condensed matter physics. As the fundamental QG entities are not gravitational ones – at least not in the usual sense of (quantized) continuum metric and matter fields – spacetime needs to be recovered starting from their collective behaviour. Thus, many QG approaches are borrowing tools from other branches of theoretical physics, especially those connected to the study of quantum many-body systems.

A recent proposal to account for the putative emergence of spacetime from an arguably non-spatiotemporal regime is the functionalist strategy. Drawing from the works by Knox (2013, 2017, 2019), Lam \& Wüthrich (2018) have argued that a certain understanding of functionalism, which portrays spacetime as {\it functionally} realized by an underlying, more fundamental set of degrees of freedom, is the path to explore to account for the emergence of spacetime. On their view, the functionalist strategy is structured in two steps. First, one has to pinpoint what is the functional role of spacetime, and then identify what plays this specific functional role in the given formal context. 

This paper aims to explore the viability of the functionalist strategy within the so-called levels of spacetime emergence (as introduced by Oriti, 2021 and successively refined and exemplified within a specific quantum gravity formalism in Oriti, 2023) a theory of QG is expected to deal with. In particular, the goal will be threefold. First, it will be outlined that the functionalist strategy may represent a heuristically valuable tool whenever a notion of emergence is at stake, in that it may provide useful guidelines for both technical and conceptual purposes. Second, it will be argued that, as one proceeds throughout the levels of spacetime emergence, the heuristic role of functionalism becomes less and less significant. While it may be useful for some goals, such as defining the appropriate mediated quantities that need to be recovered, at least approximately, in the classical, continuum case, it is not for others, such as the analysis of the different phases of the underlying QG quanta or the notion of universality envisaged by certain approaches to QG. Third, as the discussion proceeds, it will become more and more evident that functionalism here defined has much more to do with a broadly reductionist strategy than with the original goal promoted within the philosophy of mind. While its compatibility with reductionism makes room for the applicability of functionalism in physical applications in general, and quantum gravity in particular, it also weakens somehow its distinctive or novel input. As borders get blurred, so (we believe) does the import of the functionalist strategy. 

The paper is structured in the following way. The second section provides a general survey on the functionalist strategy, with a particular focus on its application to the case of spacetime emergence. Some general, critical points will be raised on the usefulness of the functionalist strategy in this last context. In the third section, moving from the taxonomy offered by Oriti (2023) on the levels of spacetime emergence, the functionalist strategy will be put to test. The fourth section focuses on a specific case-study, namely Tensorial Group Field Theory (TGFT). Finally, the conclusions will outline the pros and cons of this theoretical perspective when it comes to account for the emergence of spacetime within QG in general, and (T)GFT in particular. 

\section{Spacetime functionalism}

In the philosophy of mind, functionalism is the position according to which what makes something painful, joyful and so on is not contingent upon the internal composition of a mental state (or its \lq ontological basis\rq), but depends solely on its functions, and thus the role it occupies in the cognitive system to which something painful, joyful and so on belongs. More specifically, functionalism takes a mental state to be defined by its causal connection with sensory stimulations, other mental states, and behaviour (Levin, 2023).\footnote{Notably, in the context of QG, a broader notion of “causal relations” is to be expected. On this, see Baron \& Le Bihan (2024).} Typically associated with functionalism is thus the view that the same mental state can be realized by physically heterogeneous entities (Levin, 2023) – that mental state being multiply-realizable. On this perspective, functionalism is presented as a form of non-reductive physicalism. And this, remarkably, is the bare minimum we can require from the emergence of spacetime as realized within QG projects.\footnote{With this we simply refer to the fact that the emergence of spacetime, as described within QG projects, is meant to follow from a physicalist basis, namely discarding the possibility of there being some kind of dualist interpretation of the relationship between the non-spatiotemporal and the spatiotemporal regimes. Moreover, this represents the bare minimum requirement, as we in general expect the emergence of spacetime to be at least in principle reducible to the underlying, microphysical basis. In a sense, this bare minimum is connatural to the endeavour of theoretical physics; it does not exclude the possibility of going beyond physicalism from a more general perspective, but such a step would imply going beyond a physics-based explanation.} However, as we will try to clarify, it may still be insufficient, in the sense that (a certain type of) reduction can hardly be avoided. Historically, the first pioneering works on functionalism were Putnam (1960, 1967), Lewis (1966, 1972), Block \& Fodor (1972), and Fodor (1974).
 
Within the context of spacetime, the functionalist strategy has followed two main paths, respectively introduced by Chalmers (2012) and Knox (2013). On this see, e.g., Lorenzetti (2024). Chalmers’ functionalism represents a version of the so-called Canberra plan as a way to articulate the relationship between “problematic” and “unproblematic” terms figuring in physical theories: reconstructing spatiotemporal properties means identifying (to be read as “to reduce them to”) what plays the appropriate role for spacetime. On this understanding, functionalism is a way to unveil the reduction of spacetime to its nomic role. According to Knox, who draws from the dynamical interpretation of GR endorsed by Brown (2006), an entity can be identified with spacetime as long as it plays the appropriate spatiotemporal role, namely to describe “the structure of the inertial frames” and the associated coordinate systems (2014, p. 15).\footnote{Here we do not necessarily imply our agreement with this characterization of spacetime, and limit ourselves to an outline of the functionalist strategy. In the following, we will clarify what we take spacetime to require for its identification.} Given that, in the context of GR, it is the metric field which is in charge of accounting for that structure, it follows that, at that level, the metric field should be identified with spacetime. 

The difference between these two strands of (spatiotemporal) functionalism is that they respectively focus inter and intra-theoretically -- where, for inter and intra theoretically we respectively mean the emphasis on a single theory (say, the theory of special relativity or GR) or on the relationship between two (or more) theories (such as GR and a candidate theory of QG). In Knox \& Wallace's (2023) parlance, these two forms are termed causal-role and constitutive functionalism. The latter implies that if, at a certain level, there is something that satisfies a specific functional role, an identification, at that level, between that something and the functional role is valid (regardless of the possibility of defining a more fundamental theory in a reductionist sense). Constitutive functionalism is thus analytically true. For example, the fact that at a certain level a specific assemblage performs the table function is completely independent of any fundamental theory that may eventually explain why this is the case, thus allowing an identification, at that level, between that assemblage and the table itself. Differently, causal-role functionalism puts the emphasis on the reduction process, whereby higher-level entities/properties get mapped, via a realization including bridge laws, to the lower-level entities/properties. 

For what matters in the present discussion, let us note in passing that the legacy of functionalism can be ascertained even within the sole level of GR – regardless of a more fundamental QG theory – as the functionalist lens can be adopted to explore what, among the various (mathematical) structures defined within a single theory, should be identified with spacetime (this is indeed what in the following discussion will correspond to what we term ‘level -1’ of spacetime emergence, in the GR context). Therefore, we concur with Knox \& Wallace that the functional approach can be applied even intra-theoretically, as one interprets, within a single theory, which structures or observables or mathematical objects display the functions of interest, such as spatial distance and temporal duration. At the same time, if there are reasons to believe that a certain theory (say, GR) is not fundamental, and one aims at reconstructing it, the functionalist strategy can prove useful to set the minimal criteria and the key elements and steps for such a recovery, thus guiding the reduction process. It is in this second case that functionalism may display a practical role in directing the search towards a more fundamental theory. This is why, from now onwards, we will concentrate on functionalism as a way to articulate the (arguably reductive) relationship between lower-level and higher-level theories.\footnote{On this, see also Butterfield \& Gomes (2023).} 

For instance, Lam and W\"uthrich (2018) have recently argued that spacetime functionalism is the way to proceed to account for the emergence of spacetime, namely to recover it from a non-spatiotemporal environment. In their own words, the sole task of functionalism is to explain how the collective behavior of the underlying degrees of freedom appears, in the appropriate regime, spatiotemporal “in all relevant and empirically testable ways.” (p. 44)

Their project, which goes under the general scope of the Canberra plan,\footnote{To be more precise, spacetime functionalism is formulated as a sort of \lq Canberra plan with reversed explanatory arrow\rq, dubbed “Geneva plan”, insofar as the reduction process is reversed: the non-spatiotemporal QG degrees of freedom are warranted via their identification with spatiotemporal ones (Huggett \& W\"uthrich, {\it forthcoming}, Chapter 2).} is then structured in two steps. First, one has to pinpoint the functional role of spacetime, and then one has to identify what plays this specific functional role. In some more detail, following Kim (2005), functional reduction is implemented in the following way:

\begin{enumerate}

\item {\bf (FR1)}: The higher-level entities/properties/states to be reduced are ‘functionalized’; i.e., one specifies the causal roles that identify them.
\item {\bf (FR2)}: An explanation is given of how the lower-level entities/properties/states fill this functional role.

\end{enumerate}

\noindent When these two steps are made explicit, the higher-level entities/properties/states get realized by lower-level ones. In the context of spacetime emergence, this strategy reads:

\begin{enumerate}

\item {\bf (SF1)}: Spacetime entities/properties/states are functionalized by specifying their roles, i.e. how they (concur to) define spatiotemporal features such as spacetime localization, dimensionality, intervals, etc., in relation to other entities/properties/states in the theory.

\item {\bf (SF2)}: An explanation is given of how the fundamental entities/properties/states, postulated by the QG theory under inspection, fulfil these roles, possibly only in some approximation or sector of the theory itself.

\end{enumerate}

\noindent As within the philosophy of mind, a key assumption is that the same higher-level spatiotemporal entities/properties/states can be implemented, viz. realized, by different (kinds of) lower-level entities/properties/states. This is why the functionalist strategy within spacetime emergence is meant to justify a notion of multiple-realizability.\footnote{For a recent discussion on functionalism (and associated challenges) as applied in the context of spacetime emergence see, e.g., Le Bihan (2019).} Note that, as we understand functionalism in this context, two notions of multiple-realizability are to be considered. The first concerns the possible recovery of spacetime from different approaches to QG (such as models in causal set theory, loop quantum gravity, and string theory). The second has to do with the various instantiations, within a single approach to QG, that are expected to allow for such a recovery (such as different microscopic models, within a given QG formalism, with the same, in key aspects, coarse-grained or effective dynamics).



Keeping this in the background, we believe spacetime functionalism should be put to test on a case-by-case basis, to evaluate its tenability within different theoretical frameworks. It is precisely in this spirit that Lam and W\"uthrich (2018) have shown how this strategy can be implemented in the context of causal set theory and loop quantum gravity by indicating how to recover spacetime functions such as spatiotemporal localization, spatial distance, temporal duration, topology, and similar aspects of the geometry of spacetime. With respect to this analysis, in the following we will emphasize the important (we argue, necessary) role played by the relational reconstruction of spacetime observables, and the associated link with the functionalist strategy. Indeed, the relational framework applies also to the approaches discussed by Lam and W\"uthrich (2018), i.e. causal set theory and loop quantum gravity, although this point was not addressed there, the focus being only on how to deal with the discreteness and the quantum nature of the fundamental entities. 

A final aspect to mention, as we will get back to this point in the next section, is that Lam and W\"uthrich argue the functionalist strategy represents a departure from the building-blocks prospect, according to which spacetime is composed of underlying, non-spatiotemporal, entities (p. 51). Their critical target is the primitive ontologists' view (such as the one advocated by Esfeld, 2014, or in its most radical version Esfeld, 2020), as paradigmatically implemented in the context of quantum mechanics, according to which all there is ultimately corresponds to a set of local beables evolving in a spatiotemporal background. We will try to clarify that, despite the unavailability of local beables interpreted in this way, the functionalist strategy is actually neutral with respect to the possibility of spacetime being composed (in the mereological sense) of something arguably non-spatiotemporal. In fact, our analysis of the TGFT case, and our own terminology there, actually leans toward such ‘building blocks’ perspective.\footnote{On this aspect, see also Baron \& Le Bihan, 2022.}

\subsection{Spacetime functionalism: some concluding remarks}

Before getting into the more technical part, let us close with some further general comments on the functionalist strategy as applied in the context of spacetime emergence.

Though a fully developed and complete (let alone corroborated by observations) QG theory is currently missing, whatever that theory will be it will most likely need to assign the proper functional role to spacetime. At the very least, this can be inferred by the fact that our best theory of gravitational phenomena, i.e. GR, does it: as we presume GR (or some other field-theoretic modification of it) to be recovered in some approximation, so we believe spacetime, as described within GR, to be similarly retrieved. In this sense, the functionalist strategy seems to imply that spacetime as identified (functionally) in GR and QFT, and just like GR and QFT themselves, is “secured” with respect to the fundamental ontology.\footnote{On this, see also the relevance of functionalism for the notion of empirical (in)coherence as discussed by Huggett \& W\"uthrich, 2013, and Oriti, 2014.} 

Another interesting point concerning spacetime functionalism, which emerges, although not explicitly, in Huggett \& W\"uthrich (2021), is the one connected to the empirical assessment of a QG theory and the associated notion of grounding. The latter is generally interpreted as a non-causal metaphysical relation, or, as Fine (2012) prefers to call it, an operation, that couples either facts or entities in which the explanans and the explanandum are connected via some constitutive form of determination. On this view, what grounds is taken as more fundamental than what is grounded. In the present context, the significance of functional reduction is inversed, in the sense that the non-spatiotemporal entities of QG get vindicated via their identification with spatiotemporal ones, at least in the sense that it is at the level of spatiotemporal entities that one will find (indirect) empirical confirmation of them. This means that the theoretical (or troublesome, as they prefer to call them) terms figuring in QG are placed on a firm ontological footing only if they are able, through the “aggregate operator” (namely, the one that encodes the procedure, including dynamical aspects, that allows for the recovery of spacetime), to reproduce the spacetime role. In other words, what is more fundamental is grounded, at least epistemically, in something less fundamental. Better, as there are various forms of grounding, in this theoretical scenario a more fundamental theory is spatiotemporally grounded in (or by means of) a less fundamental one. This represents a somewhat unusual instantiation of grounding, as its directionality and that of fundamentality go in opposite ways, thus providing novel interesting cause of reflection on the debate concerning restricted {\it vis-à-vis} absolute fundamentalism. We point out, however, that this form of grounding, based on epistemic considerations, is in fact rather standard for physical theories positing new entities while being arguably more fundamental than established ones; in a way, one could argue, this apparently ‘inverted’ relation is constitutive of scientific progress broadly conceived. 


A major worry concerns what type of functionalism is at stake. For, broadly speaking, functionalism is the position according to which what makes something painful, joyful and so on is not contingent upon the internal composition of a mental state, but depends solely on its functions, and thus on the role it plays in the cognitive system to which something painful, joyful and so on belongs. This means that the same function can be played by different systems, that function being thus multiply-realizable. 
The issue becomes: In what sense the notion of spacetime can be multiply-realized once a certain approach to QG eventually supersedes the others (so that we are not just in a situation of theoretical underdetermination)? Surely, functionalism allows for a hypothetical scenario in which the same functions responsible for basic metrical and topological properties as prescribed by GR get realized by different (kinds of) fundamental entities (but see the following section for a challenge to this point), as postulated by different formalisms. At the same time, whenever one picks a specific approach to QG, it may be less clear in what sense this multiple-realizability is meant to hold. However, the type of relation between a certain microscopic configuration and a macroscopic, (putatively) emergent one would be closer to a thermodynamic or hydrodynamic analogue, in which different microscopic configurations (of the same type) can produce the same macroscopic configuration (and dynamics). This is clear from a physics point of view. But it is debatable to what extent such a characterization of what a macroscopic, thermodynamic configuration is can be cast in genuine functional terms. More specifically, for a given QG formalism/framework, it may well be that the same macroscopic, emergent physics, including spacetime physics, is produced by many different microscopic models, just like one can have the same hydrodynamics (in many relevant aspects) from many different atomic field theories, or the same critical behaviour from many different microscopic systems. The general point is that one has multiple-realizability whenever there is some degree of universality of the relevant physics. This form of multiple-realizability, which is achieved in case spacetime emergence possesses some degree of universality in the given QG framework (yet only in some regime/approximation/subset of states) is the one typically encountered in thermodynamics and hydrodynamics. Therefore, as long as the functionalist strategy does not provide an explanation for why such universality condition is meant to hold, and how the explicit mathematical reconstruction is performed,
its import remains rather pending. 

On a similar vein, Hemmo \& Shenker (2020) have criticized functionalism as a form of genuine, non-reductive physicalism. The starting point, as in the context of the debate on emergence, is that the seeming autonomy of Special-Sciences (such as chemistry, biology and psychology) entities and laws seem to require a revision of strict reductive, type-identity physicalism. The possible strategies to account for this observation are then dualism and non-reductive physicalism. Historically, the fascination of functionalism precisely relies upon its putative ability to endorse the physicalist strategy while displaying a more complex, non-reductive, type of relation between lower-level and higher-level entities, states, and laws. Hemmo \& Shenker' s argument is that the genuine multiple-realizability which is compatible with non-reductive approaches, such as functionalism, implies a form of token dualism, namely the fact that in every token producing this form of multiple-realizability there figure non-physical elements (p. 1). As a consequence, functionalism is deemed to be an inconsistent proposal: it either boils down to a disguised version of reductive physicalism, or it corresponds, {\it pace} its advocates' intent, to a form of dualism. This is why they conclude that non-reductive physicalism, which implies a combination of physicalism (of the tokens) and non-reducibility (of the kinds), is inconsistent.

For the functionalist strategy to survive this critique, it seems necessary to distinguish, in analogy with the discourse on emergence, between a weak and a strong version. On this perspective, we would qualify the first as compatible with at least in-principle reduction to the underlying basis (as discussed in Lam \& W\"uthrich 2018, Butterfield \& Gomes, 2023), while the second as requiring some form of in-principle irreducibility of the higher-level entities/properties/states with respect to the lower-level entities/properties/states. On the one hand, it is debatable how much weak functionalism would qualify as a form of genuine functionalism, in that it dispenses with the initial goal of providing a non-reductive physicalist explanation of higher-level entities/properties/states (though we would maintain that it remains non-trivial  enough to represent a potentially useful conceptual tool).\footnote{Let us emphasize here that the functionalist program, as adopted in the context of spacetime emergence, is not meant to be a form of non-reductive physicalism. This arises from the consideration that the type of emergence stemming from QG projects is generally expected to be compatible with at least some form of reduction. Our dissatisfaction precisely addresses this point: if (as we concur) the emergence of spacetime can be cast in a broadly reductionist framework, the import of functionalism gets severely undermined.} On the other hand, strong functionalism is arguably incompatible with spacetime emergence as described by QG projects, as it is incompatible with the reductionist stance.

\section{Levels of spacetime emergence and the functionalist strategy}

In this section the objective is to put the functionalist strategy to test by focusing on the levels of spacetime emergence as generically encountered in QG programmes, with a final focus on (T)GFTs as a case study. We will explore the viability of this strategy and provide some comments on its usefulness. As a preliminary remark, from GR we expect physics to be built using the notions of space and time, which thus represent the conceptual building blocks for our understanding of physics and are taken as physically robust, namely, as physical entities. At the same time, as already emphasized, there are reasons to think that such notions of space and time may require a drastic revision within approaches to QG. 

Attention will be drawn to the disappearance and re-emergence of space and time as involving several aspects, with associated epistemological, methodological and metaphysical implications (in addition, of course, to the physical and mathematical ones). In some more detail, this emergence will be explored through the conceptual lens of the functionalist strategy, by considering the different levels of spacetime emergence, and what role this strategy is meant to play at each step. 

\subsection{Level -1 of spacetime emergence}

As the general slogan of spacetime functionalism here discussed reads “spacetime is as spacetime does” (Lam \& W\"uthrich, 2018), it is first and foremost necessary to clarify what one is usually referring to when speaking about spacetime. As a starting point, suppose then, without dwelling into the debates concerning the meaning of diffeomorphism invariance or the long-standing wrangle among relationists and substantivalists,\footnote{On this topic, see, e.g., Pooley (2013) and Stachel (2014).} that spacetime just corresponds, as a working definition, to what GR tells us it is, i.e. to the quantities that, in GR, have a spatiotemporal understanding. 

What does GR require, then, for space and time?\footnote{For a discussion on the relationship between the geometrical {\it vs} dynamical interpretation of GR and spacetime functionalism, see Knox (2017).} Generally speaking, GR corresponds to a dynamical theory of continuum fields defined on a (differentiable) manifold. All there is are thus fields and the associated manifold. However, diffeomorphism invariance also implies that the theory presents some spurious, redundant structure that does not correspond to anything physically identifiable with spacetime. More specifically, we maintain that manifold points do not actually correspond to physical events, and that directions on the manifold are not spatiotemporal directions, unless instantiated by dynamical fields. What is physically relevant, and mathematically non-redundant in GR are thus relations between dynamical fields. This implies, among other things, that no absolute notion of space and time is viable. 

On this perspective, to get a notion of spatiotemporal localization, one needs to apply the relational strategy (Rovelli, 1991, 2002, Marolf, 1995, Dittrich, 2006), namely to 1. select, according to accepted criteria, some degrees of freedom among the list of dynamical fields available in GR to play the role of clocks and rods, and then 2. express all the other degrees of freedom as functions of those picked out as clocks and rods, to provide them with a spatiotemporal characterization.

While there is no consensus among QG scholars on the need for the relational strategy (by the way, no issue in QG has a consensual solution among scholars, at present), it is the most agreed upon one, at least in principle. In the present context we take it as a working notion of spatiotemporal localization that will need to be recovered throughout the various steps of spacetime emergence. On this view, GR already displays a formulation based on mathematical structures that are not already spatiotemporal in a physical sense, thus redundant structures from which spacetime needs to emerge in a more physical characterization. This can be identified as {\bf Level -1 of spacetime emergence} (Oriti, 2023):

\begin{quoting}
\noindent {\bf Level -1 of spacetime emergence}: space and time get identified as relations among dynamical fields, including the metric field. In particular, one picks out the appropriate internal degrees of freedom, viz. matter fields, to play the role of approximate rods and clocks with respect to which the location and evolution of the other degrees of freedom get ascertained.  
\end{quoting}

\noindent What is the import of the functionalist strategy at this first level of spacetime emergence? How can this level be articulated as implementing a functionalist strategy? Following the steps of the Kim's template, this implies to further specify the following:

\begin{enumerate}

\item [] {\bf SF1${\bf_{-1}}$}: Space and time get identified as specific relations among selected dynamical fields (including metric field).\footnote{More precisely, diffeo-invariant observables with a local spacetime interpretation are functions of the metric field and of the values of appropriately chosen dynamical fields used as clock and rods. That is, physical reference frames, i.e. clock a rods, {\it define} space and time, and thus localization and evolution of other dynamical fields, while the role of the metric field is to provide such quantities with an extension, i.e. a geometric characterization. While these relational observables are diffeomorphism invariant by construction, they depend (by definition) on the chosen {\it physical} frame and its physical properties (dynamical equations, energy-momentum tensor and, in the quantum domain, quantum features). In general, this dependence is not negligible and affects the predictions of the theory (in contrast with coordinate frames or choices of gauges in classical GR). The extent of such dependence is contingent upon the specific system chosen as frame, the specific observable being considered, and the regime of approximation in which one is working. It is hard to give a complete and general characterization of such effects and of the transformations relating physical frames, when they exist. This remains an open issue (see Giesel \& Thiemann 2015, Goeller, Hoehn, Kirklin 2022.)} 

\item [] {\bf SF2${\bf_{-1}}$}: Identify internal appropriate degrees of freedom, e.g. matter fields, and use them as (approximate) clocks and rods to parametrize the evolution and location of other degrees of freedom.

\end{enumerate}

\noindent This means that, at level -1 of spacetime emergence, the functionalist strategy corresponds, in fact, to the relational strategy; or, better, that the relational strategy can be seen as a functionalist one. The successive step, however, is to spell out how much functionalist the relational strategy turns out, or needs, to be: simply collapsing the relational strategy to a functional realization of spatiotemporal localization seems to provide neither ontological nor explanatory advantage. Put differently, if one wants to identify, at this level of spacetime emergence, the functional strategy with the relational one, it would be appropriate to justify such a choice on the grounds of some tangible advantage. What should that advantage be? How can functionalism facilitate the technical application or conceptual analysis of the relational strategy? In particular, the multiple-realizability implied by the functionalist perspective is not entirely apparent nor central in the relational strategy, and it is not even clear whether we should expect it, on physical grounds. 

To exemplify the issue at stake, suppose that our model of spacetime and gravity includes $N$ matter fields, and that we could consider two different sets of 4 matter fields to play the role of a physical frame, i.e. one clock and three rods, to parametrize the spatiotemporal evolution of the remaining degrees of freedom. What are the differences between the respective spatiotemporal descriptions of (gravitational) physics obtained in the two different parameterizations? This is, in general, a difficult and rather open issue.\footnote{For a thorough discussion on how to apply the relational strategy in the context of a specific formalism (namely TGFT) to QG see, e.g., Marchetti \& Oriti (2021).} A first point to notice is that, even if both choices are admissible, they may result in degenerate or bizarre spatiotemporal forms, unless some restrictions on the chosen physical frames are imposed. Second, and most importantly, properly physical reference frames provided by realistic matter fields behave very differently from coordinate frames, as routinely used in GR. Coordinate frames are expected to be obtained in the limiting, highly idealised case in which the (fields defining a) physical frame (i)
 is minimally coupled, so to influence the other dynamical entities very little; (ii) it carries very little energy-momentum, otherwise one gets too much gravitational back reaction; (iii) it is very close to classical, so to avoid excessive quantum fluctuations, at least in its observable properties used for spatiotemporal localization. In the fully idealised case, it ends up behaving like some specific coordinate frame (depending on the dynamical equations it satisfies), thus as an ideal clock (for its timelike component) and rods (for its spacelike components). Third, to the extent in which the reference system behaves as an ideal reference frame, and only in this limit, writing everything in terms of it is equivalent to writing GR in a particular gauge. This means that two ideal choices correspond to choosing two different gauges. 
 
In this picture, when two physical frames are not ideal, the relationship between the two corresponding spatiotemporal representations is neither obvious nor clear, and we may not expect, in general, that they provide the same notion of spatiotemporal localization. In the approximation in which all distinguishing physical properties of the frame are negligible, one expects the system to behave like an ideal set of clock and rods, thus like a specific coordinate frame. As a consequence, the relation between two such idealized frames is given by a diffeomorphism and does not affect physical predictions. We may also not expect that there exists a (large) set of spatiotemporal quantities which remain invariant, when going from one to the other of the two parametrizations. The usual notion of general covariance and the usual set of invariant spatiotemporal observables should instead be expected to be recovered in the idealised limit of ideal clock and rods, exactly because choosing one set or another amounts to a coordinate or gauge choice. It is only in this idealised limit that we are sure to expect, therefore, a large degree of multiple-realizability of spacetime descriptions, as constitutive of the functionalist strategy, also in the relational setting.

\subsection{Level 0 of spacetime emergence}

The next level of spacetime emergence centers on the quantum to classical transition, which is  necessary to recover spacetime as encoded in GR (or other classical gravitational theory), in the sense of level -1, starting from a more fundamental quantum description. The assumption here is that the more fundamental quantum description is obtained by turning the classical fields, including the metric, and the spatiotemporal observables constructed from it, into quantum objects, without necessarily introducing new kinds of dynamical entities. 

Although this is the more conservative approach to quantum gravity and quantum spacetime, it results already in a radical reassessment of what spacetime is at the fundamental level. All spacetime notions, such as geometric quantities like duration, curvature, volumes, as well as localization in space and time, become subject to quantum effects, such as uncertainty, superpositions, discretization. This revision calls for a reconceptualization of foundational notions such as causal relations and temporal order. There are no sharp spatiotemporal or geometric notions, and indeed no value-definiteness, but only approximately-defined and probabilistically-determined quantities, some of which will moreover acquire discrete spectra, thus forcing us also to drop our continuum intuition for spacetime. The relational strategy will need to take into account all of this, and thus also the quantum properties of the relational frames themselves, i.e. of our chosen clocks and rods (for a recent discussion see, e.g., Giacomini, 2021). 
Therefore, an additional difficulty arises from the fact that all the hard issues connected to the foundations and interpretations of quantum mechanics need to be reframed within an environment in which also the spatiotemporal structure is subjected to quantum effects. In summary:

\begin{quoting}
\noindent {\bf Level 0 of spacetime emergence}: any quantum theory of gravity will need to provide an account of how to recover classical spatiotemporal features from quantum spatiotemporal ones.
\end{quoting}

\noindent Whether this quantum to classical transition at the formal and conceptual level, or classicalization process, can be put in correspondence with specific physical processes (and which ones) needs to be evaluated on a case-by-case analysis. What is sure is that, provided there is a process responsible for this classicalization, the notion of process at stake will be utterly non-temporal (see Oriti, 2021, 2023).

Before turning the attention to how the functionalist strategy gets articulated within this level of spacetime emergence, let us add a preliminary consideration. From now on, we maintain (SF1) fixed and explain at each step how (SF2) is implemented. This is appropriate if one takes, as a basic definition for space and time, the one arising within GR through the relational strategy. Moving from the steps of the functionalist strategy, at level (0) of spacetime emergence, (SF1) and (SF2) get implemented as:

\begin{enumerate}

\item [] {\bf SF1${\bf_{-1}}$}: Space and time get identified as specific relations among selected dynamical fields (including metric field). 

\item [] {\bf SF2${\bf_{0}}$}: Explain how to go from quantum (characterized by uncertainty, superpositions, interference, entanglement,
discretization,…) to classical (characterized by geometry, causal structure, localization, duration,…) spacetime. In particular, clarify the quantum features of relational geometric (and spatiotemporal) observables and their approximate suppression in specific regimes.

\end{enumerate}
The framework is considerably more challenging at the conceptual level, and one could imagine being very demanding also at the technical level in a complete theory of \lq quantum GR\rq . Still, the tools to explore the quantum-to-classical transition would be presumably some adaptation of the ones used in standard quantum theory, including coherent states, decoherence, etc, which are in fact extensively applied, say, in canonical loop quantum gravity. 

\noindent This is indeed the sort of spacetime emergence that has been dealt with from a functionalist perspective in the loop quantum gravity context in Lam \& Wuthrich (2018). Again, how much the functionalist strategy is, in fact, useful {\it per se}, i.e. to what extent it helps clarify or solve the many issues related to the quantum-to-classical transition for a quantum spacetime, is debatable. Since these are entrenched inevitably with the foundational issues of quantum mechanics, one could also ask what are the explanatory virtues of the functionalist strategy to help solve them, for their own significance or as part of solving the issues coming with the more general QG context. 
We do not see, at the moment, which specific advantages are brought by the functional framing of spacetime emergence, as we defined it here.

\subsection{Level 1 of spacetime emergence}

Whether spacetime emergence requires further levels of conceptual analysis, associated with further levels of formal developments, depends on the perspective one takes on the quantum gravity problem, and the associated formalism. 


QG may be seen as the result of the straightforward quantization of GR (or other classical theory), based on the same fundamental entities (fields), which ‘just’ have to be turned {\it quantum} (together with all geometric observables etc). 
In this case, there is no further level to be considered beyond the two discussed above -- which, by themselves, are already rich enough of difficult and interesting issues to be tackled.

However, a number of approaches to QG take a more radical view. 
On this perspective, QG may be based on some other type of quantum entities, different from quantized fields, that collectively, in the appropriate regime, recover continuum fields as encountered (after a further classical approximation) in GR. From this standpoint, the spatiotemporal universe corresponds to a peculiar quantum many-body system -- a system with associated novel, (even) less intuitive \lq atomic structures\rq, whose macroscopic description only makes use of notions like classical or quantum fields, as well as their spatiotemporally characterized (quantum) relations.\footnote{As we see in the next section, this is the case, for example, of (T)GFTs, where the underlying structures are depicted as quantum simplices (in $d=4$, tetrahedra) with an associated Fock space (like in quantum many-body physics).} We refer to this more radical context as corresponding to {\bf Level 1 of spacetime emergence}. 

Contrary to level 0, and the straightforward ‘QG $=$ quantum GR’ perspective -- where it is assumed that the fundamental QG ontology is the same as the one of classical GR -- level 1 assumes the appearence of a new, behaviorally (and ontologically) dissimilar, set of (quantum) degrees of freedom. This implies that the very notion of gravitational field (and in fact any other field) needs to appropriately, and approximately, emerge from the collective behaviour of this underlying set of new degrees of freedom. 

As such, level 1 of spacetime emergence implies a further degree of non-spatiotemporality. On this perspective, the issue is not anymore (only) the one concerning the relationship between quantum and associated classical systems. Rather, a new type of approximation is required -- one we could intuitively refer to as that from discrete to continuum entities (in analogy with the one relating atoms to fluids).\footnote{The specific form that this transition takes depends on the specific formalism considered and the nature of the assumed fundamental degrees of freedom, and the precise ways in which they differ from continuum (quantum) fields.} Even though the two transitions -- namely, quantum to classical and discrete to continuum -- can technically be implemented together in some cases, they represent two conceptually and physically distinct limits and, in general, they should not be conflated to one another, but taken separately. 

An important aspect to be noted is that how these postulated structures are to be interpreted varies, in different QG approaches, from an ontological standpoint. In some more detail, the issue at stake is: Are these more fundamental structures physical, or do they simply represent a useful, mathematical tool that is adopted, for instance, for regularization purposes only? To provide an example, this is how one could (and often does) interpret lattice structures featuring in many QG formalisms.
Though technically this does not necessarily bring about big differences -- as in both cases one needs to apply similar coarse-graining and limiting procedures\footnote{ Let us note in passing that it is from this level onwards that the role of renormalization group methods (and associated explanatory import) becomes relevant.} -- still the interpretational discrepancy is crucial from a conceptual standpoint, for it leads to different issues and insights, and it may also produce different mathematical/physical results because it may lead to differences in the way the same technical strategies are implemented or what they are expected to produce. From a conceptual standpoint, these two interpretations are extremely dissimilar. In the first case, the theory gestures towards the genuine existence of novel, behaviourally and ontologically, unalike entities, whereas in the second one is really not leaving level 0 of spacetime emergence. In the former context, therefore, one has to deal with a further level:

\begin{quoting}
\noindent {\bf Level 1 of spacetime emergence}: a theory of QG should account for the transition (distinct yet often intertwined with the quantum to classical transition), from discrete to continuum structures, or from a non-spatiotemporal set of entities to the spatiotemporal ones emerging from their collective quantum dynamics, via adequate coarse-graining and limiting procedures. 
\end{quoting}

\noindent Turning to the functionalist strategy, a preliminary remark concerns again the ontological status of the more fundamental entities postulated by the various approaches to QG which, as already emphasized, is not the same. In particular:

\begin{enumerate}

\item {\bf Fundamental entities as mere technical tools}: according to certain approaches to QG, such as causal dynamical triangulations, tensor models, quantum Regge calculus, the postulated entities do not display an ontological robustness. They are useful, technical tools.

\item {\bf Fundamental entities as ontologically robust}: according to other approaches to QG, such as loop quantum gravity\footnote{The case of loop quantum gravity, in fact, is a bit peculiar because one can easily find, in the literature and in the community, also the viewpoint that the theory corresponds to a straightforward quantization of GR, and that its states are just states of the quantized gravitational field, as defined via connections and triads.}, causal set theory, tensorial group field theories, these fundamental entities are portrayed as physically salient, and their underlying behaviour is granted ontological robustness. 

 \end{enumerate}
 
 \noindent For the functionalist strategy, these two opposing understandings of the postulated entities result in divergent consequences. 
 Indeed, what does it mean for an entity to play the spacetime role if that entity is, strictly speaking, a formal tool? 

Only those approaches that accord physical robustness to the underlying entities can apply the strategy (and in fact have to) at a deeper level of spacetimee emergence. In other words, also from the point of view of functionalism, for those approaches that treat them as a mere regularization tool only the first two steps, namely level -1 and level 0, are available. 

Suppose to consider the set of approaches to QG according physical robustness to the underlying entities. How should the functionalist strategy be implemented at this level? The programme is structured as follows:

\begin{enumerate}

\item [] {\bf SF1${\bf_{-1}}$}: Space and time get identified as specific relations among selected dynamical fields (including metric field).

\item [] {\bf SF2${\bf_{1}}$}: Explain how to move from the microscopic theory of the hypothesized “fundamental” entities, via coarse-graining and limiting procedures, to the one focused on their collective effects. Specifically, identify how relational spacetime physics and observables are reconstructed (approximately) from the fundamental non-spatiotemporal ones, and what role they play in the more fundamental context.

\end{enumerate}

\noindent 
Importantly, at this stage, the role of functionalism is connected to the debate concerning the notion of universality, understood as implying that different systems, with different microscopic features, eventuate in (some of) the same macroscopic behaviour. Also, at this level of spacetime emergence, the explanatory role of renormalization group (and mean field) methods becomes central in the conceptual analysis. Can the functionalist strategy then offer us new insights into these two aspects? In particular, can it tell us anything about their causal {\it vs} non-causal, mathematical {\it vs} non-mathematical, reductionist {\it vs} non-reductionist nature? So far, the benefit of connoting the renormalization group framework in functional terms remains uncertain, in our view. Moreover, some authors argue that the phenomenon of universality -- and thus the associated philosophical concept of multiple-realizability -- envisaged by renormalization group methods, can be cast in reductionist terms (see, e.g., Franklin, 2018). Therefore, if spacetime functionalism is to be promoted in its “strong” version, the critique pointed out by Hemmo \& Shenker, already discussed in general terms in the previous section, applies; if instead spacetime functionalism is to be defended in its “weak” form, the problem remains of clarifying how much functionalist such a proposal actually is, with respect to its original intention.  
The point, already mentioned, is that the functionalist program, as implemented in the context of spacetime emergence, is not only compatible but actually conflating with a reductionist program and associated challenges, which it does not obviously contribute to solve or evade.

\subsection{Level 2 of spacetime emergence}

Given a quantum many-body system of non spatiotemporal entities, there is in general no unique continuum (or collective, macroscopic) limit. Rather, depending on the values of parameters, couplings, external potentials, the continuum limit can lead to different phases, that are, technically speaking, different systems with different properties and different associated effective descriptions. 

Taking GR as a valid approximate description of our world, we know that at least one of these continuum phases should be geometric and spatiotemporal. This is the one in which it should be possible to offer, in the appropriate context and upon further classical approximation, an effective reconstruction of continuum spacetime and geometry (Oriti, 2023). However, not all continuum phases will be geometric, and admit a spatiotemporal reconstruction. 

If the same set of entities, in the same type of limits, can give rise to very different continuum physics, some of which being spatiotemporal, some of which being non-spatiotemporal, one needs to face deeper ontological issues concerning the robustness of those concepts of space and time we take as the inescapable underpinning of physical reality. We also need to analyse the precise differences between different phases, and the conceptual subtleties that could explain their existence at the end of the limiting procedures. The general lesson one needs to take at face value is that the universe may be even less spatiotemporal, than originally expected, and more complex, in relation to the nature and emergence of spatiotemporal notions, then we had realized at the previous levels. We are at:

\begin{quoting}
\noindent {\bf Level 2 of spacetime emergence}: a QG formalism that grants physical robustness to the underlying constituents needs to provide an analysis of the phase diagram in which they can organize themselves in the continuum/collective limit, and of the individual features and differences of the continuum phases. This comes together with a reassessment of the metaphysical and epistemological issues prompted by the even more radically non-spatiotemporal nature of the same constituents.
\end{quoting}

\noindent Level 2 of spacetime emergence requires considering the conceptual import of another kind of non-biunivocal correspondence, which in a sense corresponds to the inverse of universality: many continuum (spatiotemporal as well as non-spatiotemporal) phases, appearing at the ‘less-fundamental’ level of description, can be the product of the same underlying, more fundamental entities. Let us note in passing that this ‘inverse universality’ provides further insights on the relationship between reduction and multiple-realizability. At this level, the application of the functionalist strategy reads:

\begin{enumerate}

\item [] {\bf SF1${\bf_{-1}}$}: Space and time get identified as specific relations among selected dynamical fields (including metric field).

\item [] {\bf SF2${\bf_{2}}$}: Offer an analysis of the phase diagram obtained from the quantum dynamics of spacetime constituents, taking into account that there is no one-to-one relation between microscopic states and macroscopic phases, and that not all the phases allow for (even approximate) spacetime reconstruction. In doing so, identify the key differences that allow for the reconstruction of relational spacetime observables and their dynamics in one phase and not in others.

\end{enumerate}

\noindent Once more, at level 2 of spacetime emergence, the task for the functionalist strategy is to contribute to solving the novel technical, methodological as well as ontological issues that one encounters due to the multiplicity of continuum phases and their associated, non-spatiotemporal nature.
Again, as things stand, adding the functionalist twist does not seem to present any novel insights, from an ontological or explanatory perspective. More specifically, it seems that functionalism so conceived merely amounts to a rephrasing, with no added explanatory power or much conceptual clarification, of the various steps to be taken, within any QG programme, in the reduction process from the non-spatiotemporal to the spatiotemporal physics.  


 \subsection{Level 3 of spacetime emergence}
 
Once one accepts that among the different possible continuum phases, in which the underlying set of non-spatiotemporal entities can organize, only some will approximately correspond to the geometric, spatiotemporal physics of classical GR (or other gravitational theory), and that both the fundamental entities and all their phases are physically salient, then also the associated phase transitions get accorded potential physical meaning. Of particular interest are of course the phase transitions between a non-spatiotemporal and a spatiotemporal phase. As the transition concerns the passage from an arguably non-spatiotemporal to a spatiotemporal phase, it has been termed {\it geometrogenesis}.

What sort of physics is associated with this type of phase transition? And, correspondingly, what sort of physical signature of the transition should be looked for in the emergent (relational) spatiotemporal dynamics (within the spatiotemporal phase)?  The issue at stake is that while in theoretical physics you typically tune your analysis based on external parameters (playing the role of an external agent) or driven by a set of dynamical degrees of freedom not included in your model (as in theoretical cosmology when dealing with phase transitions along the history of the universe), here we have two sets of challenges. First, it is less clear what could tune or drive the running of the parameters, in clear physical terms; second, it is difficult to understand what it means to cross phases, because that crossing cannot be understood as a temporal process, by definition. 

A more general point becomes thus how to ‘move across’ the phase diagram, and what is the role of the renormalization group flow in this context, i.e. in a timeless environment. Can a dynamical interpretation of the renormalization group flow be available, despite the absence of a temporal notion of evolution? 

Answering such questions may require achieving a relational understanding of the RG flow, at least within the spatiotemporal phase, where it may eventually acquire a temporal characterization. One would then need to inquire what the same relational reformulation entails, if performed outside of the spatiotemporal phase.
Indeed, the only way to “move” across the phase diagram without relying on an external agent or physical environment is to look at the RG flow and assume that the coupling constants change with the “scale” of relevant phenomena, thus driving flow of effective dynamics. On this perspective, if we want to interpret geometrogenesis as a “proto-temporal” process, then we need to apply (again) the relational strategy, namely to relate the renormalization group flow parameter (scale) to those internal degrees of freedom of the system that play a role as a relational clock, in the spatiotemporal phase.

These are only some of the additional issues encountered when investigating the (conceptual and technical aspects of the) phase transitions in the quantum gravity system. They correspond to:

\begin{quoting}
\noindent {\bf Level 3 of spacetime emergence}: A QG theory that grants physical robustness to the underlying non-spatiotemporal entities, to their continuum phases and the associated phase transitions, should provide an (mathematical, physical and conceptual) analysis of such phase transitions. In particular, it needs to account for and explain the key features of the transition between non-spatiotemporal and spatiotemporal phases.
\end{quoting} 

\noindent At this level, the application of the functionalist strategy reads:

\begin{enumerate}

\item [] {\bf SF1${\bf_{-1}}$}: Space and time get identified as specific relations among selected dynamical fields (including metric field). 

\item [] {\bf SF2${\bf_{3}}$}: Provide an interpretation of how the observables encoding (relationally) spacetime physics acquire their role and properties across quantum gravity phases and in particular at criticality, and how they are affected by the RG flow within different phases, in particular the spatiotemporal one. This may require or lead to a dynamical interpretation of the renormalization group flow itself via an upgrade of the relational strategy, allowing then to describe geometrogenesis as a physical process, and implying a notion of “temporal evolution” in the theory space/phase diagram. 
\end{enumerate}

\noindent At this level, the functionalist strategy would then suggest, if not correspond to, what we dub a \lq relational strategy reloaded\rq. We say “reloaded”, because the relational strategy here needed would take place in an environment that displays way less ‘spatiotemporally intuitive’ structure than classical GR, so that defining relational evolution becomes even more challenging, and also because we are simply not aware of a relational formulation of the renormalization group itself (but see some steps in Baldazzi et al., 2021). Pointing in this research direction is definitely an interesting outcome of the adoption of a functionalist perspective, thus in favor of its usefulness. Having said this, the same suggestion can be motivated from different perspectives, and even accepting it, it leaves the actual work of realizing concretely and to solve the other associated conceptual puzzles untouched. In addition, collapsing the relational to the functionalist strategy presents the same problems discussed at the level -1 of spacetime emergence, made more severe by the fact that we are now in an even more spaceless and timeless conceptual environment.

\section{Levels of spacetime emergence: the case of (T)GFT}
This section is meant to provide a sketchy exemplification of how the levels of spacetime emergence can be implemented concretely within one specific formalism, namely (T)GFT. Here the example focuses on temporal evolution, as defined via the relational strategy within GR, but the analysis could be easily extended to spatial localization. We also focus, within the general TGFT framework, on quantum geometric models with a richer set of data and motivated by simplicial quantum gravity as well as canonical loop quantum gravity (these often go under the label GFT, emphasizing the group-theoretic data and construction).

In these models, one has a quantum many-body system whose basic entities are quantized simplicial structures, which in the case of interest -- namely, that of $d=4$ -- correspond to quantum tetrahedra (with a straightforward discrete geometric interpretation at the individual level, but promoted to the quantum domain). There is a Hilbert (Fock) space associated to generic ensembles of these simplicial structures, often formulated in a second-quantized language of creation/annihilation operators and corresponding fields (with the quantum tetrahedra being the associated quanta). The gluing among these simplicial structures, which encodes discrete space connectivity, is associated with their mutual entanglement. 

Just like one has a discrete geometry associated to such \lq atoms of space\rq, one can also include additional data interpreted as discretized matter fields associated to the same simplicial structures. This is a key ingredient for a proper physical significance of these models, of course, but also for applying (at some level of approximation) the relational strategy to reconstruct a (quantum) spacetime. For example, one focuses on (T)GFT models for simplicial geometry coupled with free, mass-less, minimally coupled scalar fields. 

Since the fundamental degrees of freedom are not quantized continuum fields, we are clearly facing the issues of level 1 of spacetime emergence. The question is then how to recover, from this fundamental structures, an effective continuum spacetime physics via their collective dynamics. 
Dealing with an interacting quantum many-body system, we should expect that this collective dynamics will give rise to several inequivalent phases (level 2) and associated phase transitions (level 3), with all the anticipated conceptual issues. To proceed with the recovery of an effective continuum spacetime, a number of hypotheses need to be made.

The first hypothesis is that the geometric phase corresponds to a condensed phase of the atoms of spacetime, rather than to any generic continuum phase. This restricts the search and implies some implicit assumptions on the issues of level 2. The second hypothesis is that we can adopt the hydrodynamic-like approximation to recover the spatiotemporal notions encountered in quantum (and then classical) GR. This amounts to a specific strategy and perspective for tackling the issues of level 1. Put it differently, the suggestion is that the spatiotemporal universe corresponds to a quantum gravity condensate or a peculiar quantum fluid, described by an associated condensate wavefunction (or mean field, or, equivalently, density and phase of the fluid) as the key collective variable to use for the extraction of spatiotemporal information.

All the techniques that, in quantum many-body physics and the theory of quantum fluids, can be used to obtain an effective hydrodynamics for the condensate, starting from the full partition function (encoding the complete quantum dynamics of the atoms of space), can in principle be applied to the quantum gravity system in the TGFT formulation. They are the formal tools to solve the issues of level 1.

A simple approximation method, much used in TGFTs, particularly in the cosmological sector for the extraction of spacetime physics, is the mean-field approximation.
Mean field hydrodynamics corresponds to operating with simple (T)GFT states, namely coherent states of the field operator, which (although fully captured by a single condensate wavefunction on the domain of data for a single quantum tetrahedron) encode a collective description of a large number (potentially infinite) of quantum tetrahedra. 

Having obtained the hydrodynamic equations and thus being able, in principle, to compute any TGFT observable, in the same approximation, the task becomes to compute those which admit a spatiotemporal interpretation.  They will necessarily amount to coarse-grained, averaged quantities, in line with level 1 of spacetime emergence. This also means extracting an effective spatiotemporal dynamics, in the language of field theory and geometry, from the more abstract TGFT hydrodynamics.

The key fact used to proceed is that the TGFT mean field is a function over a domain isomorphic to the space of homogeneous spatial geometries, namely the minisuperspace in gravitational physics (which is also the domain of the wavefunction of quantum cosmology, i.e., a space of field values). Because of this, the associated mean field equations of motion correspond to a non-linear extension of the dynamics for quantum cosmology -- where the linear dynamical operator is the Wheeler-DeWitt one of a canonical quantization. 
Despite this extension, and the associated interpretational differences, one could now proceed in a similar fashion as in quantum cosmology (or canonical quantum gravity) to compute geometric observables and their effective dynamics, and then, in a final step, their classical approximation. That is, one is now equipped to tackle the issues of level 0 of spacetime emergence too.
To recast everything so far discussed in the spatiotemporal parlance, one introduces a physical relational frame, and reformulates the effective hydrodynamics in terms of relational, spatiotemporal observables. We are still in a fully quantum domain, from the point of view of geometric observables, thus at level 0 of spacetime emergence, assuming we are focusing already on collective, averaged quantities (thus having moved there from level 1). 

This geometric reformulation can be done in several ways (see the TGFT cosmology literature, e.g. Pithis \& Sakellariadou, 2019).
One procedure involves considering  quantum condensate states in which the effective scalar field data (the mean field counterpart of the fundamental discrete scalar field data included in the TGFT model) behaves appropriately to be used as the relational clock, in particular by imposing a sort of semiclassicality condition on them. 

In this way, one can obtain relational observables -- such as the number operator or the universe volume -- at a given clock time, as function of the scalar field (thus, as prescribed by level -1 of spacetime emergence), via an analysis of the expectation value on the selected condensed states (thus, indeed, as hydrodynamic averages, in the spirit of level 1). But such observables are still highly quantum, in principle (as they could have large quantum fluctuations), and thus we have not yet dealt with the issues of level 0 of spacetime emergence. Looking at the relational volume at a given clock time, you can extract its temporal evolution equation and then show that it reproduces the classical GR dynamics, with suppressed fluctuations,  for large values of the chosen clock, thus solving (at least for this observable and special class of states) the issues of level 0 of spacetime emergence. At earlier time the physics remains quantum and, for example, one gets a quantum bounce instead of the Big Bang singularity for the universe evolution. 

This indicates a possible path from the most fundamental level of non-spatiotemporal description to classical GR. Many and crucial approximations and restrictions are necessary parts of this path. We have recovered a geometric picture for specific observables and states, in a particular hydrodynamic approximation, in a particular condensed phase of the fundamental degrees of freedom, and for a specific class of TGFT models. But this, of course, is not the whole story. Indeed, we should find the new physics beyond classical GR (and new conceptual insights) when analysing in detail the quantum nature of observables and dynamics (level 0 of spacetime emergence), the breakdown of the hydrodynamic regime at sufficiently small scales (going towards level 1 from level 0), and the critical regime separating the spatiotemporal condensate phase from the other phases and exploring them too (level 2 and level 3 of spacetime emergence). 

\section{Conclusions}

Any future theory of quantum gravity will have to recover GR and usual spacetime physics in the appropriate regime of approximation. From a heuristic point of view, the functionalist strategy clarifies what the minimal requirements for recovering spacetime are. It also provides ways to ‘protect’ spacetime, from an ontological and epistemological perspective, in some sense, from its complete disappearance when moving to more fundamental quantum gravity formalisms in which it does not feature as such. 
At the same time, some comments must be added. 

First, working physicists may favour different options for the functional spacetime roles to be granted, than those implied by the relational strategy implies (for instance, at the level -1 of spacetime emergence). Although this point may be circumvented by adopting a form of minimal empiricism, and specifying the functional roles of spacetime may be sufficient to grant empirical validation, GR comes equipped with a more complex apparatus (world view), which goes well beyond the pure description of observed correlations. For instance, one would like to understand where the equivalence principle comes from, when it is valid, whether the interpretation of gravity as geometry of spacetime is the only option, if the fields are an approximation and of what, and these are not, strictly speaking, just empirical questions. It would be important, then, to clarify how the functionalist strategy can help solve such issues. 

Another aspect one needs to carefully consider is what is the ontological status of the entities postulated within a certain approach to QG, in the case in which these are granted physical \lq reality\rq of some sort. 
The notion of existence is tied both to both observational consequences as well as explanatory value within the theory. Can the functionalist strategy be viable in those approaches in which the “realizers” are ultimately non-spatiotemporal? 

In addition, one needs to spell out what notion of multiple-realizability is involved in a functionalist framing of spacetime emergence, and at all its levels. As already emphasized, we may distinguish between strong and weak functionalism (as in the case of emergence). If the situation is the one typically targeted by multiple-realizability, namely that classical GR can be recovered by different theoretical approaches (such as models in string theory, GFT or causal set theory), it follows that functionalism corresponds to a mere reformulation of the involved terms and issues. In doing so, it does not appears to add much to their understanding or solve ensuing problems or give an explanation for how it is the case that these different models recover the same spatiotemporal structure. 
If the situation is more nuanced, as in fact we expect, namely if we admit that each quantum gravity model will only partly recover the spacetime structure of GR, only approximately and in some corner of the theory (as in fact one should await in any reduction-compatible emergence context, we would maintain), it looks as if we are dismantling the main historical motivation for introducing functionalism. The situation would be entirely compatible with reduction (and in some sense just another set of clothes for it), possibly useful, but of limited novel import. In addition, if one considers the notion of inverse universality, as implied by the possible existence of many inequivalent collective phases on the basis of the same fundamental entities, this implies that the link between the geometric phase and the underlying, pre-geometric one is not necessary, thus highlighting the novelty of the notions of space and time which arise in the geometric phase but also the radically non-spatiotemporal nature of the fundamental constituents. Also on this point, functionalism seems to have little to say.

All the above general comments and issues find a concrete exemplification in the case of TGFT models of spacetime emergence.
In particular , it exemplifies both universality and inverse universality in spacetime emergence, in the same way in which condensed matter physics does. 
In GFTs, at the more fundamental level there is a many-body quantum system that is treated via an hydrodynamic approximation, to see how it behaves at an effective level and if this can be recast in spatiotemporal, geometric terms. One sees a many-to-one relation at work. This is what happens, for instance, when in GFT cosmology one evaluates the volume of the universe by considering the contributions of the various atoms of space: many microstates, which are physically inequivalent, will give rise to that averaged volume and, in the hydrodynamic approximation, to its relational evolution. And in the same coarse-graining procedure (and for the same geometric observable), many details that differentiate different specific GFT models of the fundamental quantum dynamics of atoms of space become irrelevant; indeed, a qualitatively similar effective dynamics for the universe has been obtained starting from different microscopic TGFT models. As it takes place in a condensate phase, the recovery of spacetime envisaged by this approach can be put in correspondence with the emergence of novel phases of matter in condensed matter physics. It represents a form of weak emergence, whereby spacetime as described by GR is dependent upon, yet provides boundary constraints to, the underlying GFT quanta. But one sees also the one-to-many inverse universality relation, since the condensate phase is just one of the phase that can be obtained from the same fundamental quantum entities (quantum tetrahedra described within a given TGFT model).

Further, it is not very clear in what sense the emergence of spacetime could not be taken in compositional terms, namely as arising out of the collective behaviour of an ensemble of non-spatiotemporal entities. This is precisely what we are looking for in TGFT and we have managed to (partially) realize, thus obtaining an explicit counterexample to the contrary claim.  
So, either Lam \& W\"uthrich's argument according to which spacetime functionalism would lead away from the “building blocks” heuristics is not sound, or GFT (and kindred approaches) cannot be cast in functional terms.\footnote{However, this second option should be ruled out, based on what we argued throughout the various levels of spacetime emergence, at least within a specific understanding of functionalism.} A third option would be to clarify that QG projects require a non-standard form of mereology that is compatible with spacetime functionalism, yet distant from the “building blocks” heuristics.

Finally, QG projects force practitioners, and philosophers too, to reason on a whole new class of heavy conceptual problems to be solved within the various levels of spacetime emergence, among which we have mentioned only a small part. For example, what does it mean to apply a relational strategy? Correspondingly, what notion of observable is at stake? How can a notion of process be introduced into a formalism like equilibrium statistical mechanics where, technically, no notion of time is available? How should this situation be interpreted in the absence of an external observer? What can be said about the ontology of the fundamental degrees of freedom, if one takes into account all the different phases of the QG quanta? 

It would be interesting to see whether the functionalist strategy may be further developed to provide novel insights on these heavy conceptual issues. For the time being, as we have tried to show, the concrete conceptual import of functionalism as applied to the emergence of spacetime remains uncertain.

\end{document}